# IMPROVEMENT OF ATTENTION IN SUBJECTS DIAGNOSED WITH HYPERKINETIC SYNDROME USING BIOVIT SIMULATOR

**Dr. Cesar R Salas-Guerra, PhD, DBA**

*cesar.salasg@autonoma.cat*

*Doctoral Program, Philosophy Department,*

*Autonomous University of Barcelona*

*Catalunya*

*2021*

# Abstract

This study aimed to stimulate the brain's executive function through a series of tasks and rules based on dynamic perceptual stimuli using the Biotechnology Virtual Immersion Simulator (BIOVIT) and thus evaluate its usefulness to maintain and increase attention levels in subjects diagnosed with hyperkinetic syndrome. With a quantitative methodology framed in a longitudinal trend design, the cause of the exposure-outcome relationships was studied using the BIOVIT simulator. Exploratory analysis of oscillatory brain activity was measured using a graphical recording of brain electrical activity and attention levels. Data consisted of 77,566 observations from n = 18 separately studied participants. The findings established that the BIOVIT simulator maintained and increased the attention levels of the participants by 77.8%. Furthermore, the hypothesis was tested that virtual reality immersion technologies significantly affect attention levels in participants aged 8 to 12. The evidence shows that the BIOVIT simulator is an alternative to developing learning methodologies in vulnerable populations. The low implementation costs and the diversity of academic applications may allow schools in developing countries to solve this problem that afflicts thousands of children with attention deficit and hyperactivity disorder.

# 1. Introduction

Today human beings have developed new adaptive capacities based on information management through executive and cognitive functions. We can mention attention and memory as an ability to remember and carry out operations in the future (Grandi & Tirapu-Ustárroz, 2017) through structured processes such as coding, interpretation, and storage (Fombuena, 2016)

However, inadequate sensory processing regulations are identified as an additional dimension of the hyperkinetic syndrome (Navarra & Waterhouse, 2019). Hyperkinetic syndrome or attention deficit with hyperactivity (ADHD) most commonly affects the child population (Rubiales Josefina, Liliana Bakker, Diana Russo, 2014). There is substantial evidence linking hyperkinetic syndrome to various sensory processing dysfunction problems (Navarra & Waterhouse, 2019), presenting sensory modulation problems compared to others without (ADHD).

Psychostimulant agents are described in some studies as a treatment to aid in patients' quality of life; though, ecological testing with virtual reality is a very encouraging option that benefits modulation of sensory signal processing and improves its performance (Parsons et al., 2017) because immersion in virtual reality allows the development of dynamic visual stimuli by optimally encoding the sensory information received (Młynarski & Hermundstad, 2018) and minimizes the distraction of incoming stimuli (Yuste et al., 2017).

Within the paradigm that establishes virtual reality's potential to revolutionize education (Rubiales Josefina, Liliana Bakker, Diana Russo, 2014), (Marian et al., 2018), virtual reality environments may allow the user to immerse themselves in a simulation of everyday activities. This human-computer interface (HCI) requires a robust user experience (UX) design to track translational and rotational movements (Parsons et al.,

2017,) facilitating the development of dynamic perceptible stimuli. One of them is visual and auditory information. These factors produce specific effects seen in the literature as beneficial for increasing and retaining attention naturally and frequently (Marian et al., 2018).

Therefore, this study aims to test the BIOVIT simulator that allows stimulating the executive function of the brain infusion into the real world through a series of tasks and rules based on dynamic perceptual stimuli (Parsons et al., 2017) by using virtual immersion biotechnology, thus managing to maintain and increase attention levels in subjects diagnosed with the hyperkinetic syndrome.

Several functions include the processing, propagation, and synchronization of information attributed to the brain's oscillating activity (Kissinger et al., 2018) and making inferences about the environment to plan and meet goals successfully (Młynarski & Hermundstad, 2018).

## 2. Theoretical framework: information processing theory and the neurophysiology of learning

The information processing theory contributes to learning through stimuli used by attention, perception, and memory (Moos, 2015). Recent literature establishes a relationship between the functions of the oscillatory activity of the brain, such as the processing, propagation, and synchronization of information (Kissinger et al., 2018) with attention, with low-frequency oscillations being significant in the processes of perception of visual stimuli (Einstein et al., 2017).

These findings show that one of the qualities of attention is extracting the characteristics of a stimulus (Grandi & Tirapu-Ustárroz, 2017), increasing the brain activity involved in processing information and receptivity of active stimuli (Ruiz-Contreras & Cansino, 2005). Ecological testing with virtual reality is a very encouraging option that benefits sensory signal processing modulation and improves its performance (Parsons et al., 2017) because immersion in virtual reality allows the development of dynamic visual stimuli by optimally encoding the sensory information received (9) and minimizing the distraction of incoming stimuli (Yuste et al., 2017).

Therefore, new learning tools such as virtual immersion biotechnology composed of cognitive, affective, and physiological qualities would help students perceive, respond, and interact in different real-life situations (Hoffmann, Agustín Freiberg; Liporace, 2015).

Consequently, this theoretical approach demonstrates the importance of attention in storing information's coding and interpretation processes, and short-term memory develops sufficient semantic recovery resources (Luna-Lario et al., 2017). With this premise, the following hypothesis was formulated, which will be verified in the study.

H1: Virtual immersion biotechnology is beneficial for retaining and improving attention among children with hyperkinetic syndrome.

## 3. Research methodology

This studio consists of a quantitative approach and longitudinal trend design, which seeks to identify the possible cause in the exposure-result ratio using the virtual immersion biotechnology simulator (BIOVIT) and exploratory analysis of brain electrical activity and attention levels. The data will be collected and analyzed according to the ethical protocols applicable to this study. Below is explained in detail the methodology used and the protocols implemented.

### *3.1 Participants*

This study evaluated 77,566 observations from 18 students from 8 to 12 years of age diagnosed with the hyperkinetic syndrome. Recruitment was carried out incidentally by disseminating the live voice project through information exposed in the respective associations of support for relatives of patients with hyperkinetic syndrome in two schools in the north of Minas Gerais, Brazil.

Participation was voluntary, unpaid, and with the prerequisite of reading and approving the research project's information sheet and the signature of their respective informed consent. The sample was an intentional pseudorandom selected through the inclusion and exclusion criteria, the parents' respective authorities, and the individual academic institutions' ethics committee under current scientific recommendations (Hueso & Cascant, 2012). The inclusion criteria are described in appendix 1.

*3.2 Data Security*

The data generated in the study were stored in a digital repository using information encryption processes, thus complying with the policies of integrity, traceability, and adequate preservation during the time established by the systematic generation of periodic backups, which will be stored for a current period of five years (UAB, 2013).

A copy of this will be returned to the subject after the study is completed, ensuring that it is not disseminated by mistake or without care. Similarly, the program (Software) used in the patient data recording process was archived for the time established in the protocol.

*3.3 Potential research risk*

Invasive among the risks or discomfort that may arise from volunteers' participation in the study, it is imperative to establish the difference between non-invasive and invasive brain electrical activation studies. The first is an s electroencephalogram, a neurophysiological examination that uses electrodes overlaid on the scalp to record activity in the cerebral cortex.

The Second, electrocorticography, is a neurophysiological examination that uses electrodes placed directly on the brain's exposed surface to record activity in the cerebral cortex through craniotomies or in the operating room during surgery. Therefore, EEG is recognized as a safe and non-invasive method for neuroeducation applications based on electronic data generation devices.

**4. Methodological procedure**

This section describes the methodological procedure used for data collection and analysis, with a brief integrative review that includes empirical and technical literature. Therefore, it can provide a complete understanding of how the research was conducted in the field. This procedure included the following steps: 1) the brain electrical activity register, 2) virtual immersion biotechnology simulator (BIOVIT), 3) ecological validation of virtual immersion biotechnology test (BIOVIT), and 4) human computer-interface environment.

*4. 1 First phase research: The brain electrical activity register*

Within this macro-physiological data recording process, the participants were instructed to keep calm and collaborate with the monopolar electroencephalogram (Yuste et al., 2017); the specifications and characteristics of the EGG used in the study are described below: (Table 1)

This device consists of an external cranial adapter located on the left side of the forehead called in the electrodeposition nomenclature as the left hemisphere's Fp1 frontopolar region (Morillo, 2005); the data collection device has algorithms called eSense™ [1](NeuroSky, 2010) which measure attention and meditation levels.

These algorithms develop dynamic oscillation processes and spectra by adapting natural fluctuations based on each participant's trends using visual, audible, and tactile stimuli from the BIOVIT virtual immersion simulator. The eSense algorithm™ set a measurement scale for attention and meditation consisting of low performance (1-39), normal or base performance (40-60), and superior performance (61-100).

### *4.2 Second phase research: Virtual immersion biotechnology simulator (BIOVIT)*

Recent literature emphasizes the need for a new generation of tests led by functions that allow the stimulation of the executive brain system in the real world and the possibility of tasks adapting according to the study population (Pallavicini et al., 2019). The conceptual incorporation of the MET mandates test (Shallice & Burgess, 1991) with the use of a virtual immersion biotechnology simulator (BIOVIT) offered the possibility of obtaining a detailed record of the individual performance of each participant, combining the rigor and control of tests with fixed stimuli performed in the laboratory, to simulations that reflect dynamic stimuli in real-life situations (Pallavicini et al., 2019).

The BIOVIT tool allowed to stimulate the brain's executive function infusion into the real world through a series of tasks and rules based on dynamic perceptual stimuli (Parsons et al., 2017) by using virtual immersion technology, and this tool is composed of two dimensions: resource use and time use. In the test's execution, participants were asked to meet the objectives that require problem-solving and decision-making within the virtual immersion environment (Mäkinen et al., 2020).

---

[1] Algorithm patented by NeuroSky

Several previous studies have provided sufficient evidence of the efficiency of virtual reality content and video games in evaluating executive functions (Parsons et al., 2017). Recently, virtual reality content and video games have been studied separately. The user experience (UX) between the traditional spaces and the new virtual reality environments allows study participants to satisfactorily meet the objectives (Serino & Repetto, 2018). The efficiency of the design (UX) in these new environments is due to the degree of spatial tracking in the user interface that does not depend only on the relative sensory orientation or the orientation senses but on the immersive experience through the degrees of freedom in the rotation-translation movements through tracking and depth sensors. (Figure 1)

## *4.3 Third phase research: Ecological validation of virtual immersion biotechnology test (BIOVIT)*

Ecological assessment is called a "function-based approach" (Serino & Repetto, 2018), which involves direct observation of behavior that includes daily life tasks. This validation process establishes the importance of likelihood (Heart Stroke Foundation, 2019) as a replicability principle within behaviors of interest based on direct observation (Pallavicini et al., 2019).

Of the existence of outdated technologies and static stimuli related to the traditional use of laboratory tests, all this needs to include the contexts of affection and motivation required in evaluating real-world activities (Parsons, 2015). Indeed, virtual immersion technology has been identified as suitable for developing ecologically valid environments, as 3D objects in spatial computing are presented consistently and accurately (Parsons et al., 2017).

Therefore, the BIOVIT test was used as part of a new generation of video-based virtual immersion tools, which proved ready to motivate and challenge the participant, providing easy access to experimentation based on "fun" and a sense of commitment and self-efficacy, supporting the brain to acquire the new and complex skill (Serino & Repetto, 2018), (Prasad et al., 2018).

*4.4 Four phases of research: Human computer-interface environment*

Through a human-computer interface (HCI), the MET test could be adapted with a usability design that helps users perform their tasks smoothly (Hassanien & Azar, 2015). It has also made it possible to combine and develop new techniques, such as using haptic devices adapted to virtual immersion simulators and collecting brain activity simultaneously with the test's execution in real time.

Considering the MET's conceptual framework (Shallice & Burgess, 1991) within the requirements of the study, it was established the need for the participant in the virtual immersion process to use a haptic device that helps in the acquisition of skills in specific tasks of the execution of the test (Prasad et al., 2018); in addition to having proven in previous studies, its valuable contribution to learning.

The haptic device used with the virtual reality simulator was a steering wheel that will contribute to developing a level of commitment generating emotion (Mäkinen et al., 2020) due to different degrees of pressure since (Girod et al., 2016), in some studies, it has been identified that these devices used together with virtual reality simulators contribute to the proper use of the hands to perform specific tasks.

Therefore, the haptic device and the virtual environment selected for this test were the virtual simulator Real Feel Racing (Bonetti et al., 2018) which has tree-degree-of-freedom (3DoF) tracking, which tracks rotational movements based on sensors

(accelerometers, gyroscopes, and magnetometers) built into the mobile device (Pak & Maoz, 2019).

This virtual immersion simulator consists of eight racetracks, which the participant can select. This simulator consists of four models of race cars identified with different colors. The participant selects the panoramic view within the immersion environment, among which they can choose the perspective mode, foreground, background, and third plane. (Figure 2)

*4.4.1 Task and Rules Guideline*

This test was based on a guideline that was explained to each participant in their native language and included:

Task section: in this section, the participant was asked to select the track, the car color, and the panoramic view. Besides, he was asked to finish the race in the shortest time possible while avoiding accidents that could delay it. (Figure 3)

Rules section: in this section, the participant was asked not to leave the race before the time necessary to complete the evaluation, nor to exceed the time required to complete the evaluation, and finally, not remove the VR goggles before the time required to complete the evaluation. (Figure 4)

**5. Measurement research model**

The data were analyzed with the Minitab version 18.1 program for MS Windows 10, where descriptive and inferential statistics were performed, and a new measurement scale was designed. To comply with the assumptions of normality, the Anderson-Darling test was performed using the hypothesis test to search for normalcy.

*5.1 First phase data analysis: Hypothesis test for normality search*

Before performing the hypothesis test, the assumption of normality is verified:

$H_o$ Data follow a normal distribution

$H_1$ Data do not follow a normal distribution

Given the test results with *p-values* of *.005* in AD, the null hypothesis is rejected. Therefore, according to the nature of this type of data, the data collected do not follow a normal distribution. (Table 2)

*5.2 Second phase data analysis: Central trend descriptive tests*

The essential characteristics of the study data are described below. They provide a simple summary of the sample and measurements. (Table 3)

*5.3 Third phase data analysis: Creation of the Measurement Model*

The attention growth analysis model was created based on a non-stationary time series. The respective preliminary tests were carried out with the linear model, exponential growth, quadratic, and S curve; to determine which model best fits the data of this research.

The data obtained in the MAPE (mean absolute percentage error), MAD (absolute deviation of the mean), and MSD (squared deviation of the mean) tests were compared. These tests allowed us to observe that the quadratic model has a more significant data fit since the results obtained were lower than the other models, thus confirming what the literature review mentions.

The selected model includes a non-stationary time series showing the growth trend. The quadratic model estimation uses the following function:

$$Y_t = \beta_0 + \beta_1 t + \beta_2 t^2 + e_t$$

Where $Y_t$ is the value of attention in time, $\beta_0$ is the constant, $\beta_{1-2}$ the coefficients, $t$ time unit value and $e_t$ the error term. Therefore, if $\beta_1, \beta_2 > 0$, $Y_t$ observation is monotonous increasing, if $\beta_1, \beta_2 < 0$, $Y_t$ is monotonous decreasing. In the same way if $\beta_1 > 0$ y $\beta_2 < 0$, $Y_t$ is concave down, and if $\beta_1 < 0$ y $\beta_2, > Y_t$ its shape is concave upwards. Complete analysis in appendix 2.

### *5.4 Four-phase data analysis: Time measurement scale*

The selected time scale (t) was in minutes and seconds as the collection process using the BIOVIT tool was performed on a minute and second scale that was the maximum time that lasted the respective tests *t = 5.37*. (Table 4)

### *5.5 Five-phase data analysis: new measurement scale design*

#### *5.5.1 EFGA degrees of the care frequency scale*

Considering that the scale assigned by the eSense algorithm™ for attention and meditation is: low performance (*1-39*), average or base performance (*40-60*), superior performance (*61-100*), and since a scale that fits the study variables has not been found in the literature review, a scale called "Attention Grade Frequency Scale - EFGA" was designed for this study, which consists of the following function:

$$f_{biovit} = \frac{M_o}{Mape} x 100$$

Where $f$ is the value of the composite scale of attention as a function of the stimulation of the virtual immersion technology tool (BIOVIT). $M_o$ are repeated values more frequently within the observations. MAPE is the absolute percentage error medium. Thus, the higher the scale result, the better the care process.

*5.5.2 Internal validation and reliability estimation*

Cronbach's Alpha test and Pearson's Correlation were used to perform the Internal validation of the EFGA degree of attention frequency scale, seeking to evaluate the internal uniformity of the scale, as well as the strength and direction of the relationship between elements; the results yielded $α = .959$ and $r = .922$; therefore, the EFGA will be kept as a scale within this investigation. (Table 5)

*5.6 Six-phase data analysis: Mann-Whitney hypothesis test*

The previous growth trend test results presented minimum scores in the mean absolute percentage error MAPE, which validated the model's fit and precision from the growth of the repeated values more frequently ($M_o$) within the observations.

The nonparametric Mann-Whitney test was performed to validate and determine if the two groups' medians differ significantly, since the higher the attention growth, the lower the mean absolute percentage error. Therefore, the first hypothesis test was performed separately with the first fourteen participants, $n = 14$. The hypotheses of the Mann-Whitney test are described below:

$\boldsymbol{H_o}$    $n_1 - n_2 = 0$

$\boldsymbol{H_1}$    $n_1 - n_2 > 0$

Given that the p-value = .001 is less than the significance level of 0.05, we rejected the null hypothesis. We concluded that the level of care was maintained and grew considerably using the BIOVIT tool. (Table 6)

The second Mann-Whitney hypothesis test was performed, which is described below:

$\boldsymbol{H_o}$    $n_1 - n_2 = 0$

$$H_1 \quad n_1 - n_2 < 0$$

Given that the *p-value = .002* is less than the *0.05* level of significance, we reject the null hypothesis and conclude that the level of attention decreased considerably with the BIOVIT tool. (Table 7)

## 6. Discussion and Conclusions

Comparing the theoretical model and the results of the tests carried out; we can conclude that the BIOVIT tool maintained and increased the participants' attention levels by 77.8%. Consistent with the hypothesis test results, it was possible to conclude that the BIOVIT tool has a statistically significant effect on the retention and increase of attention in children aged 8 to 12 years diagnosed with hyperkinetic syndrome.

Therefore, BIOVIT is an alternative for developing biocybernetics learning methodologies in vulnerable populations. The low implementation costs and the diversity of academic applications can allow educational centers in developing countries to solve this problem that afflicts thousands of children with attention deficit and hyperactivity disorder.

Finally, due to the type of study population, which has characteristics and treatments, in addition to the financial resources that an international investigation contemplates, certain limitations were identified about longitudinal follow-up that could allow observing changes in the attentional processes of the daily life of the participant, as well as integration tests in traditional academic environments with the use of virtual immersion technology.

**Appendix 1 Inclusion and exclusion criteria**

Participants met the following inclusion criteria as follows:

(1) Residents in Montes Claros, State of Minas Gerais, Republic of Brazil.

(2) Have a clinical diagnosis of attention deficit and hyperactivity.

(3) Do not receive psychostimulant medicine within 24 hours before the test.

(4) His first language must be Portuguese.

(5) Age ranges from 8 to 12 years.

(6) Fundamental teaching schooling.

(7) Hearing and vision physical conditions should be adequate to participate in the test, using if necessary, corrective prosthetic measures, such as using glasses, hearing aids, or any other device.

(8) Sufficient ability to see, listen, and use either of its two limbs to use the virtual immersion simulator.

(9) The following factors were considered for exclusion criteria:

(10) Lack of will or inability of the participant to collaborate appropriately in the study.

(11) Any central nervous system pathology may affect cognition such as Parkinson's disease, Huntington's disease, brain tumour, hydrocephalus, progressive supranuclear epilepsy, subdural hematoma, multiple sclerosis, hand history of cerebral infarction.

(12) A major depressive episode or dysthymic disorder, according to DSM-V criteria.

(13) Unstable or clinically significant cardiovascular disease in the previous six months may impact mental abilities in the clinician's opinion.

(14) Be patient with diabetes with insulin dependence level.

(15) Any situation that, in the opinion of the principal investigator, is unsuitable for the study.

## Appendix 2 Quadratic Trend Analysis 1

| Variable | $M_o$ | Mape | EFGA | QUADRATIC TREND ANALYSIS | eSense™ |
|---|---|---|---|---|---|
| MOC1-3 | 88 | 13.53 | 6.50 | Gráfica de análisis de tendencia de MOC1-3<br>Modelo de tendencia cuadrática<br>$Yt = 89.376 - 0.000305 \times t + 0.000000 \times t^2$ | >61 Superior |
| MOC2-4 | 100 | 17.50 | 5.71 | Gráfica de análisis de tendencia de MOC2-4<br>Modelo de tendencia cuadrática<br>$Yt = 59.455 + 0.000720 \times t - 0.000000 \times t^2$ | >61 Superior |
| MOC0-3 | 100 | 17.87 | 5.61 | Gráfica de análisis de tendencia de MOC0-3<br>Modelo de tendencia cuadrática<br>$Yt = 86.088 - 0.000508 \times t + 0.000000 \times t^2$ | >61 Superior |
| MOC2-1 | 100 | 22.04 | 4.53 | Gráfica de análisis de tendencia de MOC2-1<br>Modelo de tendencia cuadrática<br>$Yt = 67.339 + 0.000805 \times t - 0.000000 \times t^2$ | >61 Superior |
| MOC1-8 | 41 | 22.22 | 1.84 | 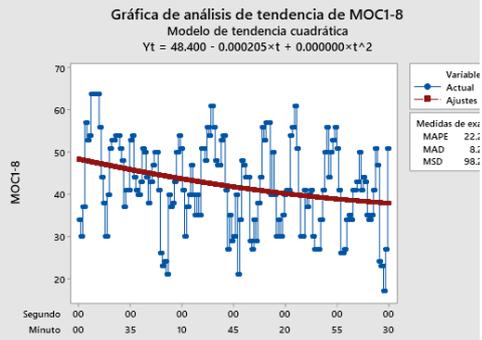 | >40 Normal |
| MOC2-5 | 63 | 24.00 | 2.62 | 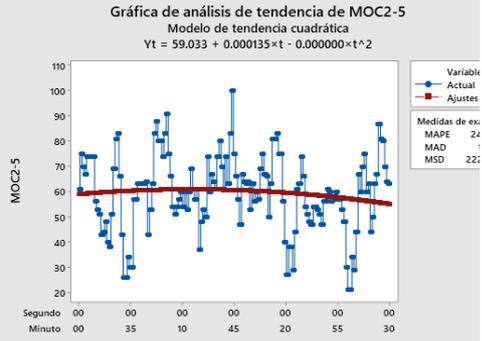 | >61 Superior |
| MOC1-5 | 51 | 25.12 | 2.03 | 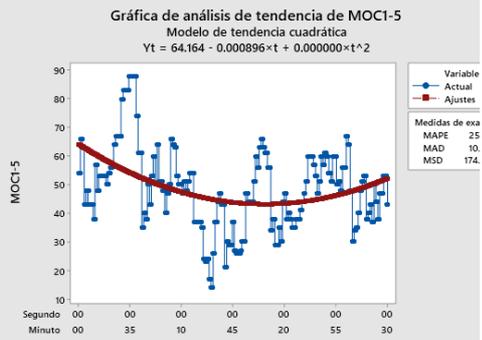 | >40 Normal |
| MOC2-3 | 50 | 25. 22 | 1.98 | 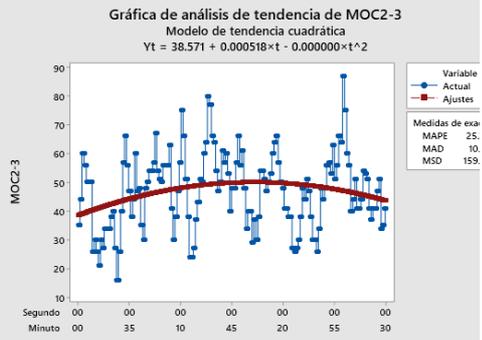 | >40 Normal |
| MOC0-4 | 61 | 29.39 | 2.07 | 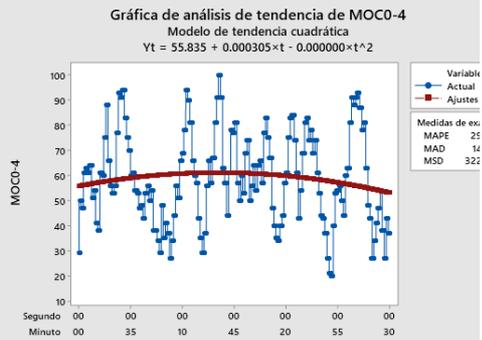 | >61 Superior |
| MOC1-1 | 47 | 32.51 | 1.44 | 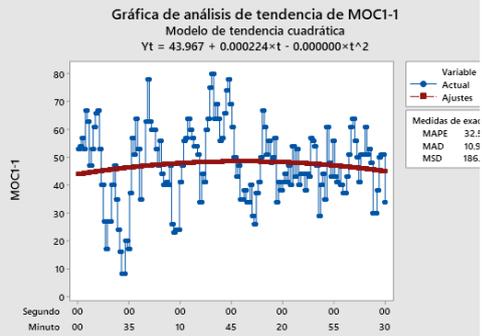 | >40 Normal |
| MOC1-7 | 57 | 34.20 | 1.66 | 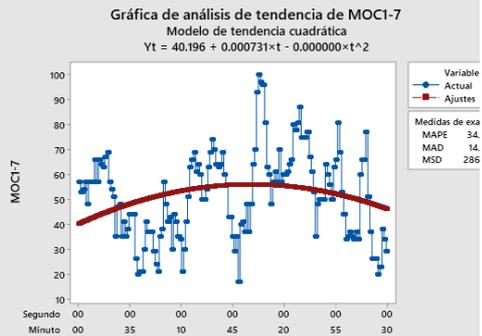 | >40 Normal |
| MOC1-2 | 37 | 35.78 | 1.03 | 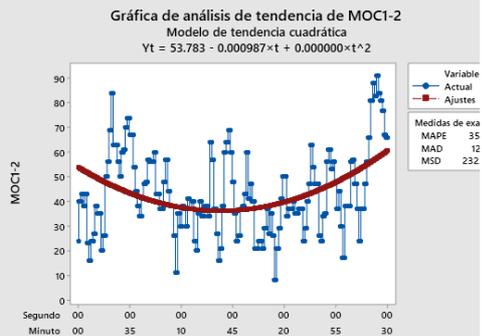 | <39 Bass |
| MOC1-6 | 53 | 36.97 | 1.43 | 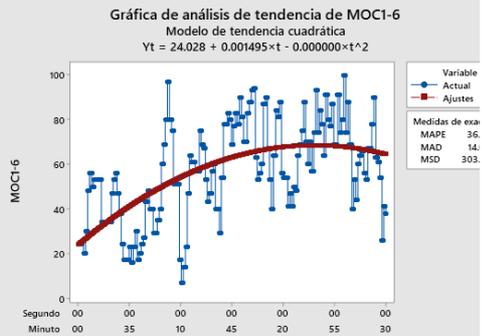 | >40 Normal |
| MOC0-2 | 53 | 48.22 | 1.09 | 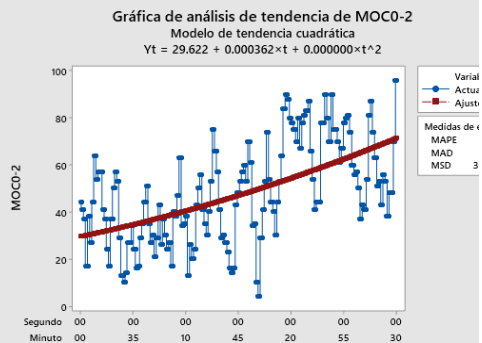 | >40 Normal |

**Quadratic Trend Analysis 2**

| Variable | $M_o$ | MAPE | EFGA | QUADRATIC TREND ANALYSIS | eSense™ |
|---|---|---|---|---|---|
| MOC1-4 | 30 | 48.28 | 0.62 | Gráfica de análisis de tendencia de MOC1-4. Modelo de tendencia cuadrática. $Yt = 42.186 + 0.000012 \times t + 0.000000 \times t^2$. MAPE 48.2, MAD 13., MSD 297. | <39 Bass |
| MOC2-2 | 47 | 54.21 | 0.86 | Gráfica de análisis de tendencia de MOC2-2. Modelo de tendencia cuadrática. $Yt = 45.040 + 0.000898 \times t - 0.000000 \times t^2$. MAPE 54., MAD 15., MSD 396. | >40 Normal |
| MOC0-1 | 41 | 73.27 | 0.55 | Gráfica de análisis de tendencia de MOC0-1. Modelo de tendencia cuadrática. $Yt = 41.827 + 0.000054 \times t + 0.000000 \times t^2$. MAPE 73., MAD 11., MSD 221. | >40 Normal |
| MOC2-6 | 30 | 77.34 | 0.38 | Gráfica de análisis de tendencia de MOC2-6. Modelo de tendencia cuadrática. $Yt = 58.223 - 0.000721 \times t + 0.000000 \times t^2$. MAPE 77., MAD 11., MSD 204. | <39 Bass |

# Appendix 3 Tables and Figures

**Table 1.** EEG Monopolar NeuroSky Specifications

| Specifications | NeuroSky EEG | Diagram |
|---|---|---|
| **Weight:** | 90g | |
| **Top sensor dimensions:** | 225mm x width:155mm x depth: 92mm | |
| **Lower sensor dimensions:** | 225mm x width:155mm x depth:165mm | |
| **Force Range:** | 30mw to 50mw - 6dBm RF max power | |
| **Frequency:** | 2,420 - 2.471GHz RF | |
| **Data Range:** | 250kbit/s RF | |
| **Range:** | 10m RF | |
| **Packet Loss:** | 5% via Wireless | |
| **Uart baudrate:** | 57,600 baud | |
| **Maximum input signal:** | 1mV pk range signal | |
| **Hardware Range Filter:** | 3Hz to 100hz | |
| **ADC Resolution:** | 12bits | |
| **Sample range:** | 512Hz | |
| **Esense calculation range:** | 1Hz | |

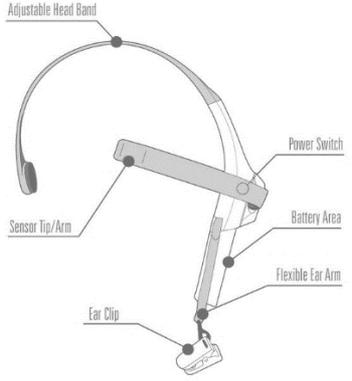

**Table 2.** Anderson-Darling Test Results

| Variables | Average | Des.Est. | Observations | And.Darling | P-value |
|---|---|---|---|---|---|
| MOC0-1 | 45.233 | 15.062 | 77566 | 300.082 | .005 |
| MOC0-2 | 48.179 | 21.226 | 77566 | 531.850 | .005 |
| MOC0-3 | 77.642 | 15.217 | 77566 | 988. 968 | .005 |
| MOC0-4 | 58.862 | 18.083 | 77566 | 405.. 306 | .005 |
| MOC1-1 | 47.218 | 13.702 | 77566 | 223.. 417 | .005 |
| MOC1-2 | 43.284 | 16.593 | 77566 | 1021. 765 | .005 |
| MOC1-3 | 78.769 | 13.137 | 77566 | 945.077 | .005 |
| MOC1-4 | 43.080 | 17.258 | 77566 | 402.820 | .005 |
| MOC1-5 | 48.602 | 14.321 | 77566 | 392. 797 | .005 |
| MOC1-6 | 56.832 | 21.716 | 77566 | 408.. 314 | .005 |
| MOC1-7 | 51.642 | 17.421 | 77566 | 396. 607 | .005 |
| MOC1-8 | 42.252 | 10.371 | 77566 | 512. 637 | .005 |
| MOC2-1 | 73.134 | 17.820 | 77566 | 556.. 770 | .005 |
| MOC2-2 | 54.353 | 20.798 | 77566 | 682.. 612 | .005 |
| MOC2-3 | 46.945 | 12.966 | 77566 | 298.004 | .005 |
| MOC2-4 | 72.466 | 14.850 | 77566 | 459.. 790 | .005 |
| MOC2-5 | 59.452 | 15.008 | 77566 | 398.853 | .005 |
| MOC2-6 | 40.626 | 16.167 | 77566 | 274.. 991 | .005 |

**Table 3.** Descriptive statistics of experimental conditions

| Variable | Observations | Average | Variance | Desv.Est. | $M_o$ |
|---|---|---|---|---|---|
| MOC0-1 | 77566 | 45.233 | 226.849 | 15.062 | 41 |
| MOC0-2 | 77566 | 48.179 | 450.533 | 21.226 | 53 |
| MOC0-3 | 77566 | 77.642 | 231.546 | 15.217 | 100 |
| MOC0-4 | 77566 | 58.862 | 327.000 | 18.083 | 61 |
| MOC1-1 | 77566 | 47.218 | 187.747 | 13.702 | 47 |
| MOC1-2 | 77566 | 43.284 | 275.318 | 16.593 | 37 |
| MOC1-3 | 77566 | 78.769 | 172.584 | 13.137 | 88 |
| MOC1-4 | 77566 | 43.080 | 297.848 | 17.258 | 30 |
| MOC1-5 | 77566 | 48.602 | 205.092 | 14.321 | 51 |
| MOC1-6 | 77566 | 56.832 | 471.580 | 21.716 | 53 |
| MOC1-7 | 77566 | 51.642 | 303.492 | 17.421 | 57 |
| MOC1-8 | 77566 | 42.252 | 107.561 | 10.371 | 41 |
| MOC2-1 | 77566 | 73.134 | 317.549 | 17.820 | 100 |
| MOC2-2 | 77566 | 54.353 | 432.568 | 20.798 | 47 |
| MOC2-3 | 77566 | 46.945 | 168.117 | 12.966 | 50 |
| MOC2-4 | 77566 | 72.466 | 220.522 | 14.850 | 100 |
| MOC2-5 | 77566 | 59.452 | 225.239 | 15.008 | 63 |
| MOC2-6 | 77566 | 40.626 | 261.385 | 16.167 | 30 |

**Table 4.** Attention growth analysis

| Variable | $M_o$ | MAPE | Mad | Msd | $BIOVIT_t$ |
|---|---|---|---|---|---|
| MOC0-1 | 41 | 73.270 | 11.678 | 221.284 | 5.14 |
| MOC0-2 | 53 | 48.216 | 14.719 | 304.994 | 5.37 |
| MOC0-3 | 100 | 17.865 | 12.529 | 222.564 | 3.52 |
| MOC0-4 | 61 | 29.386 | 14.666 | 322.506 | 3.09 |
| MOC1-1 | 47 | 32.512 | 10.998 | 186.171 | 3.13 |
| MOC1-2 | 37 | 35.782 | 12.154 | 232.832 | 3.09 |
| MOC1-3 | 88 | 13.530 | 9.600 | 139.165 | 3.17 |
| MOC1-4 | 30 | 48.280 | 13.512 | 297.418 | 5.28 |
| MOC1-5 | 51 | 25.124 | 10.484 | 174.809 | 3.26 |
| MOC1-6 | 53 | 36.967 | 14.060 | 303.367 | 3.43 |
| MOC1-7 | 57 | 34.203 | 14.258 | 286.215 | 5.09 |
| MOC1-8 | 41 | 22.224 | 8.264 | 98.226 | 4.04 |
| MOC2-1 | 100 | 22.039 | 12.723 | 269.204 | 3.36 |
| MOC2-2 | 47 | 54.210 | 15.829 | 396.084 | 3.25 |
| MOC2-3 | 50 | 25.217 | 10.177 | 159.141 | 3.25 |
| MOC2-4 | 100 | 17.500 | 11.716 | 199.137 | 3.22 |
| MOC2-5 | 63 | 24.003 | 11.611 | 222.764 | 2.32 |
| MOC2-6 | 30 | 77. 343 | 11.759 | 204.672 | 3 |

**Table 5.** Normal-superior EFGA attention scale

| Variable | $M_o$ | Mape | EFGA |
|---|---|---|---|
| MOC1-3 | 88 | 13.53 | 6.50 |
| MOC2-4 | 100 | 17.50 | 5.71 |
| MOC0-3 | 100 | 17.87 | 5.61 |
| MOC2-1 | 100 | 22.04 | 4.53 |
| MOC1-8 | 41 | 22.22 | 1.84 |
| MOC2-5 | 63 | 24.00 | 2.62 |
| MOC1-5 | 51 | 25.12 | 2.03 |
| MOC2-3 | 50 | 25.22 | 1.98 |
| MOC0-4 | 61 | 29.39 | 2.07 |
| MOC1-1 | 47 | 32.51 | 1.44 |
| MOC1-7 | 57 | 34.20 | 1.66 |

**Table 6.** Low EFGA attention scale

| Variable | $M_o$ | MAPE | EFGA |
|---|---|---|---|
| MOC2-2 | 47 | 54.21 | 0.86 |
| MOC1-4 | 30 | 48.28 | 0.62 |
| MOC0-1 | 41 | 73.27 | 0.55 |
| MOC2-6 | 30 | 77.34 | 0.38 |

**Table 7.** First Mann-Whitney hypothesis test

| Variable | $M_o$ | MAPE | N.Trust. | IC Inf. | p-value |
|---|---|---|---|---|---|
| MOC1-3 | 88 | 13.53 | 95.32% | 23.13 | .001 |
| MOC2-4 | 100 | 17.50 | 95.32% | 23.13 | .001 |
| MOC0-3 | 100 | 17.87 | 95.32% | 23.13 | .001 |
| MOC2-1 | 100 | 22.04 | 95.32% | 23.13 | .001 |
| MOC1-8 | 41 | 22.22 | 95.32% | 23.13 | .001 |
| MOC2-5 | 63 | 24.00 | 95.32% | 23.13 | .001 |
| MOC1-5 | 51 | 25.12 | 95.32% | 23.13 | .001 |
| MOC2-3 | 50 | 25.22 | 95.32% | 23.13 | .001 |
| MOC0-4 | 61 | 29.39 | 95.32% | 23.13 | .001 |
| MOC1-1 | 47 | 32.51 | 95.32% | 23.13 | .001 |
| MOC1-7 | 57 | 34.20 | 95.32% | 23.13 | .001 |
| MOC1-2 | 37 | 35.78 | 95.32% | 23.13 | .001 |
| MOC1-6 | 53 | 36.97 | 95.32% | 23.13 | .001 |
| MOC0-2 | 53 | 48.22 | 95.32% | 23.13 | .001 |

**Table 8.** Second Mann-Whitney hypothesis test

| Variable | $M_o$ | MAPE | N.Trust. | IC Inf. | p-value |
|----------|-------|-------|----------|---------|---------|
| MOC1-4   | 30    | 48.28 | 95.37%   | -7.21   | .002    |
| MOC2-2   | 47    | 54.21 | 95.37%   | -7.21   | .002    |
| MOC0-1   | 41    | 73.27 | 95.37%   | -7.21   | .002    |
| MOC2-6   | 30    | 77.34 | 95.37%   | -7.21   | .002    |

**Figure 1.** Three-degree-of-freedom (3DoF)

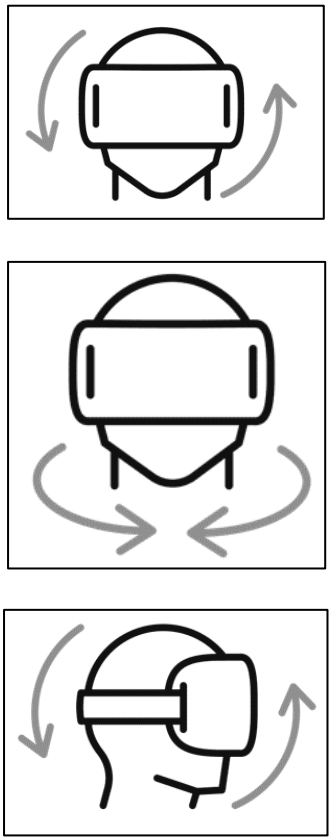

**Note:** Three-degree-of-freedom (3DoF) tracking tracks rotational movements. It is the simplest form of monitoring and relies entirely on the sensors (accelerometers, gyros, and magnetometers) built into mobile phones that use virtual reality headsets to measure movement.

**Figure 2.** Real Feel Racing Virtual Simulator

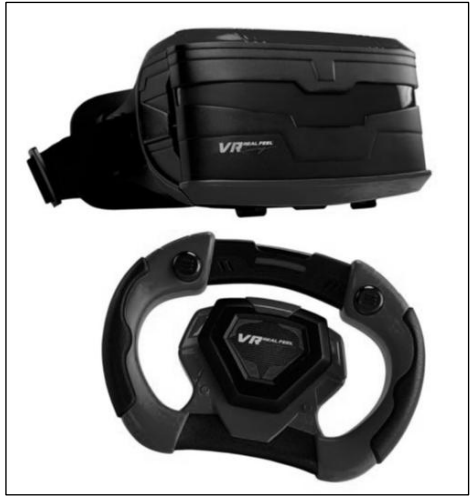

**Figure 3.** BIOVIT Task imagen

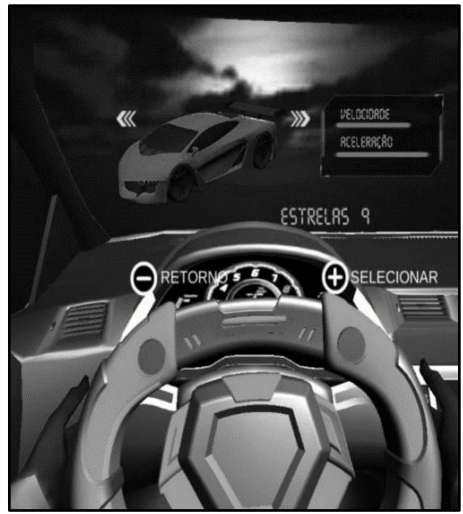

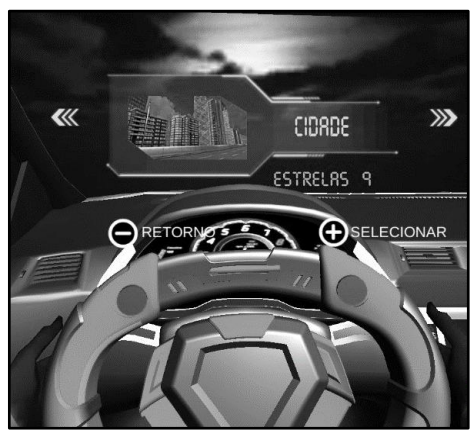

**Figure 4.** BIOVIT Rules imagen

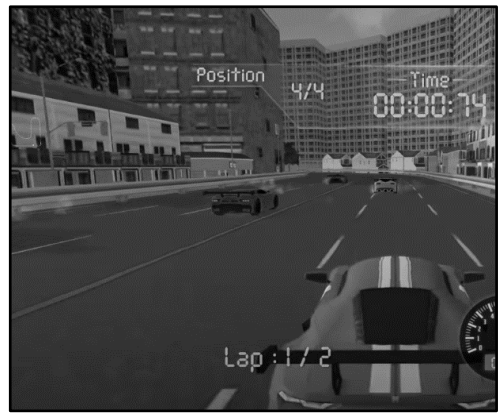

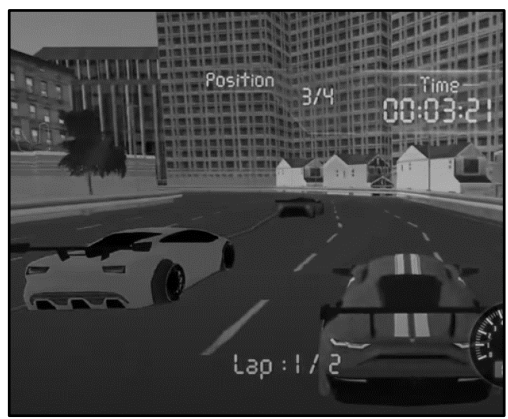